\documentclass[journal=jacsat,manuscript=article]{achemso}
\setkeys{acs}{articletitle = true} 
\usepackage{xr} 
\usepackage{longtable}
\usepackage{multirow}
\usepackage{amsmath}
\usepackage{amssymb}
\usepackage{subfigure}
\usepackage{booktabs}
\usepackage[usenames,dvipsnames]{color}
\SectionNumbersOn
\author{Oliver T.\ Unke}
\affiliation{Department of Chemistry, University of Basel, Klingelbergstrasse 80, CH-4056 Basel, Switzerland\\ Present Address: Machine Learning Group, TU Berlin, Marchstr. 23, 10587 Berlin, Germany}
\author{Debasish Koner}
\affiliation{Department of Chemistry, University of Basel, Klingelbergstrasse 80, CH-4056 Basel, Switzerland.}
\author{Sarbani Patra}
\affiliation{Department of Chemistry, University of Basel, Klingelbergstrasse 80, CH-4056 Basel, Switzerland.}
\author{Silvan K\"aser}
\affiliation{Department of Chemistry, University of Basel, Klingelbergstrasse 80, CH-4056 Basel, Switzerland.}
\author{Markus Meuwly}
\affiliation{Department of Chemistry, University of Basel, Klingelbergstrasse 80, CH-4056 Basel, Switzerland and
Department of Chemistry, Brown University, Providence, RI, USA.}
\email{m.meuwly@unibas.ch}

\date{\today}

\title{High-Dimensional Potential Energy Surfaces for Molecular
  Simulations}

\begin{document}

\begin{abstract}
An overview of computational methods to describe high-dimensional
potential energy surfaces suitable for atomistic simulations is
given. Particular emphasis is put on accuracy, computability,
transferability and extensibility of the methods discussed. They
include empirical force fields, representations based on reproducing
kernels, using permutationally invariant polynomials, and neural
network-learned representations and combinations thereof. Future
directions and potential improvements are discussed primarily from a
practical, application-oriented perspective.
\end{abstract}

\section{Introduction}

\noindent
The dynamics of molecular (i.e.\ chemical, biological and physical)
processes is governed by the underlying intermolecular
interactions. These processes can span a wide range of temporal and
spatial scales and make a characterization and the understanding of
elementary processes at an atomistic scale a formidable
task.\cite{MM.rev.sd:2017} Examples for such processes are chemical
reactions or functional motions in proteins. For typical organic
reactions the time scales are on the order of seconds whereas the
actual chemical step (i.e.\ bond breaking or bond formation) occurs on
the femtosecond time scale. In other words, during $\sim 10^{15}$
vibrational periods energy is redistributed in the system until
sufficient energy has accumulated along the preferred ``progression
coordinate'' for the reaction to occur.\cite{weber:2015} Similarly,
the biological process of ``allostery'' couples two (or multiple)
spatially separated binding sites of a protein which is used to
regulate the affinity of certain substrates to a protein, thereby
controlling metabolism.\cite{cui:2008} According to the conventional
view of allostery, a conformational change of the protein (that might
however be very small~\cite{Nussinov15}) is the source of a signal,
but other mechanisms have been proposed as well which are based
exclusively on structural dynamics\cite{cooper84}. Here, binding of a
ligand at a so-called \textit{allosteric site} increases (or
decreases) the affinity for a substrate at a distant \textit{active
  site}, and the process can span multiple time and spatial scales to
the extent of the size of the protein itself. Hence, an allosteric
protein can be viewed as a ``transistor'', and complicated feedback
networks of many such switches ultimately make up a living
cell~\cite{alon07}. As a third example, freezing and phase transitions
in water are entirely governed by intermolecular
interactions. Describing them at sufficient detail has been found
extremely challenging and a complete understanding of the phase
diagram or the structural dynamics of liquid water is still not
available.\\

\noindent
All the above situations require means to compute the total energy of
the system computationally efficiently and accurately. The most
accurate method is to solve the electronic Schr\"odinger equation for
every configuration $\vec{x}$ of the system for which energies and
forces are required. However, there are certain limitations which are
due to the computational approach {\it per se}, e.g. the speed and
efficiency of the method or due to practical aspects of quantum
chemistry such as accounting for the basis set superposition error,
the convergence of the Hartree-Fock wavefunction to the desired
electronic state for arbitrary geometries, or the choice of a suitable
active space irrespective of molecular geometry for problems with
multi-reference character, to name a few. Improvements and future
avenues for making QM-based approaches even more broadly applicable
have been recently discussed.\cite{qiang.jcp:2016} For problems that
require extensive conformational sampling or sufficient statistics
purely QM-based dynamics approaches are still impractical.\\

\noindent
A promising use of QM-based methods are mixed quantum
mechanics/molecular mechanics (QM/MM) treatments which are
particularly popular for biophysical and biochemical
applications.\cite{senn:2009} Here, the system is decomposed into a
``reactive region'' which is treated with a quantum chemical (or
semiempirical) method and an environment described by an empirical
force field. Such a decomposition considerably speeds up simulations
such that even free energy simulations in multiple dimensions can be
computed.\cite{cui:2016} One of the current open questions in such
QM/MM simulations is that of the size of the QM region required for
converged results which was recently considered for Catechol
O-Methyltransferase.\cite{kulik:2016}\\

\noindent
Other possibilities to provide energies for molecular systems are
based on empirical energy expressions, fits of reference energies to
reference data from quantum chemical calculations, representations of
the energies by kernels or by using neural networks. These methods are
the topic of the present perspective as they have shown to provide
means to follow the dynamics of molecular systems over long time
scales or to allow statistically significant sampling of the process
of interest.\\

\noindent
First, explicit representations of energy functions are
discussed. This usually requires one to choose a functional form of
the model function. Next, machine learned potential energy
surfaces are discussed. In a second part, topical applications of
these methods are presented.\\

\section{Explicit Representations}
\label{sec:explicit_representations}
Empirical force fields are one of the most seasoned concepts to
represent the total energy of a molecular system given the coordinates
$\vec{x}$ of all atoms. The general expression for an empirical FF
includes bonded $E_{\rm bonded}$ and nonbonded $E_{\rm nonbonded}$
terms.
\begin{eqnarray}
E(\bf x)&=&\sum_{\rm bonds} k_b (r-r_e)^2 + 
           \sum_{\rm angle} k_{\theta} (\theta-\theta_e)^2 \nonumber
           \sum_{\rm dihedrals} k_{\chi} (1+\cos{n \chi - \delta}) +
            \sum_{\rm impropers} k_{imp} (\phi - \phi_0)^2 \nonumber \\
        &+&\sum_{\rm nonbonded} \epsilon_{ij} 
            \left[ \left( \frac{R_{\rm min_{\rm ij}}}{r_{ij}} \right)^{12}- 
            \left ( \frac{R_{\rm min_{\rm ij}}}{r_{ij}}  \right )^{12} \right]+
            \frac{q_i q_j}{\epsilon_l r_{ij}}
\label{eq:ff}
\end{eqnarray}
Such representations can be evaluated very efficiently, the forces are
readily available and systems containing millions of atoms can be
simulated for extended time scales. On the other hand, the accuracy of
such force fields compared with high-level electronic structure
methods is very limited. Conversely, one of the noteworthy advantages
of empirical energy functions is that they can be consistently and
incrementally improved. Examples include the replacement of harmonic
potentials for chemical bonds by Morse oscillator functions or
extending the point charge electrostatics through multipolar series
expansions.\cite{ponder07multi,kramer2013,tristan2013,Pengyu2013}
Also, additional terms can be included to provide a more physically
motivated representation, such as adding a term for polarization
interactions.\cite{lopes2013}\\

\noindent
For smaller molecular systems more accurate representations are
possible. Typically, reference energies are computed from quantum
chemical calculations on a grid (regular or irregular) of molecular
geometries. These energies are then used to fit parameters in a
predetermined functional form to minimize the difference between the
reference energies and the model function.\\

\noindent
One example for such a predefined functional form are permutationally
invariant polynomials (PIPs) which have been applied to molecules with
4 to 10 atoms and to investigate diverse physico-chemical
problems.\cite{bowman.irpc:2009} Using PIPs, the permutational
symmetry arising in many molecular systems is explicitly built into
the construction of the parametrized form of the PES. The monomials
are of the form $y_{ij} = \exp{\left(-r_{ij}/a\right)}$ where the
$r_{ij}$ are atom-atom separations and $a$ is a range parameter. The
total potential is then expanded into multinomials, i.e. products of
monomials with suitable expansion coefficients. For an A$_2$B molecule
the symmetrized basis is $y_{12}^a(y_{13}^b y_{23}^c + y_{23}^b
y_{13}^c)$ which explicitly obeys permutational symmetry. A library
for constructing the necessary polynomial basis has been made publicly
available. \cite{bowman:2010}\\

\noindent
One application of PIPs includes the dissociation reaction of CH$_5^+$
to CH$_3^+$ + H$_2$ for which more than 36000
energies\cite{bowman.ch5:2006} were fitted with an accuracy of 78.1
cm$^{-1}$. With this PES the branching ratio to form HD and H$_2$ for
CH$_4$D$^+$ and CH$_5^+$, respectively, was determined. Also, the
infrared spectra of various isotopes were computed with this
PES.\cite{bowman.ch52:2006} Other applications concern a fitted energy
function for water dimer,\cite{bowman.water:2008} which became the
basis for the WHBB force field for liquid water\cite{bowman.whbb:2011}
and that for acetaldehyde.\cite{bowman:2007} For acetaldehyde roughly
135'000 energies at the CCSD(T)/cc-pVTZ level of theory were fitted to
2655 terms with order 5 in the polynomial basis and 9953 terms with
order 6 in the polynomial basis. For the relevant stationary states in
that study the difference between the reference calculations and the
fit ranges from 2 to 4.5 kcal/mol. However, the overall RMSD for all
fitted points has not been reported.\cite{bowman:2007} With this PES
the fragment population for dissociation into CH$_3$ + HCO and CH$_4$
+ CO was investigated.\\

\noindent
Another fruitful approach are double many body
expansions.\cite{dmbe:1988} These decompose the total energy of a
molecular system first into one- and several many-body terms and then
represent each of them as a sum of short- and long-range
contributions.\cite{dmbe:1988} This yields, for example, an RMSD of
0.99 kcal/mol for 3701 fitted points from electronic structure
calculations at the MRCI level of theory for CNO.\cite{varandas:2018}
As a comparison, another recent investigation of the same
system\cite{MM.cno:2018} using a reproducing kernel Hilbert space
(RKHS, see further below) representation yielded an RMSD of 0.38, 0.48
and 0.47 kcal/mol for the $^2$A$'$, $^2$A$''$ and $^4$A$''$ electronic
states using more than 10000 {\it ab initio} points for each
surface.\\

\noindent
Local interpolation has also been shown to provide a meaningful
approach. One such approach is Shepard interpolation which represents
the PES as a weighted sum of force fields, expanded around several
reference geometries.\cite{collins2002molecular,dawes:2009} Also,
recently several computational resources have been made available to
construct fully-dimensional PESs for polyatomic molecules such as
Autosurf\cite{autosurf:2019} or a repository to automatically
construct PIPs. \\

\section{Machine Learned PESs}
\label{sec:machine_learned_pess}
Machine learning (ML) methods have become increasingly popular in
recent years in order to construct PESs, or estimate other properties
of unknown compounds or
structures.\cite{rupp2012fast,montavon2013machine,hansen2013assessment,hansen2015machine}
Such approaches give computers the ability to learn patterns in data
without being explicitly programmed.\cite{samuel2000some} For PES
construction, suitable reference data are e.g.\ energy, forces, or
both, usually obtained from \textit{ab initio} methods. Contrary to
the explicit representations discussed in
section~\ref{sec:explicit_representations}, ML-based PESs are
non-parametric and not limited to a predetermined functional form.\\

\begin{figure}
\includegraphics[width=0.375\textwidth]{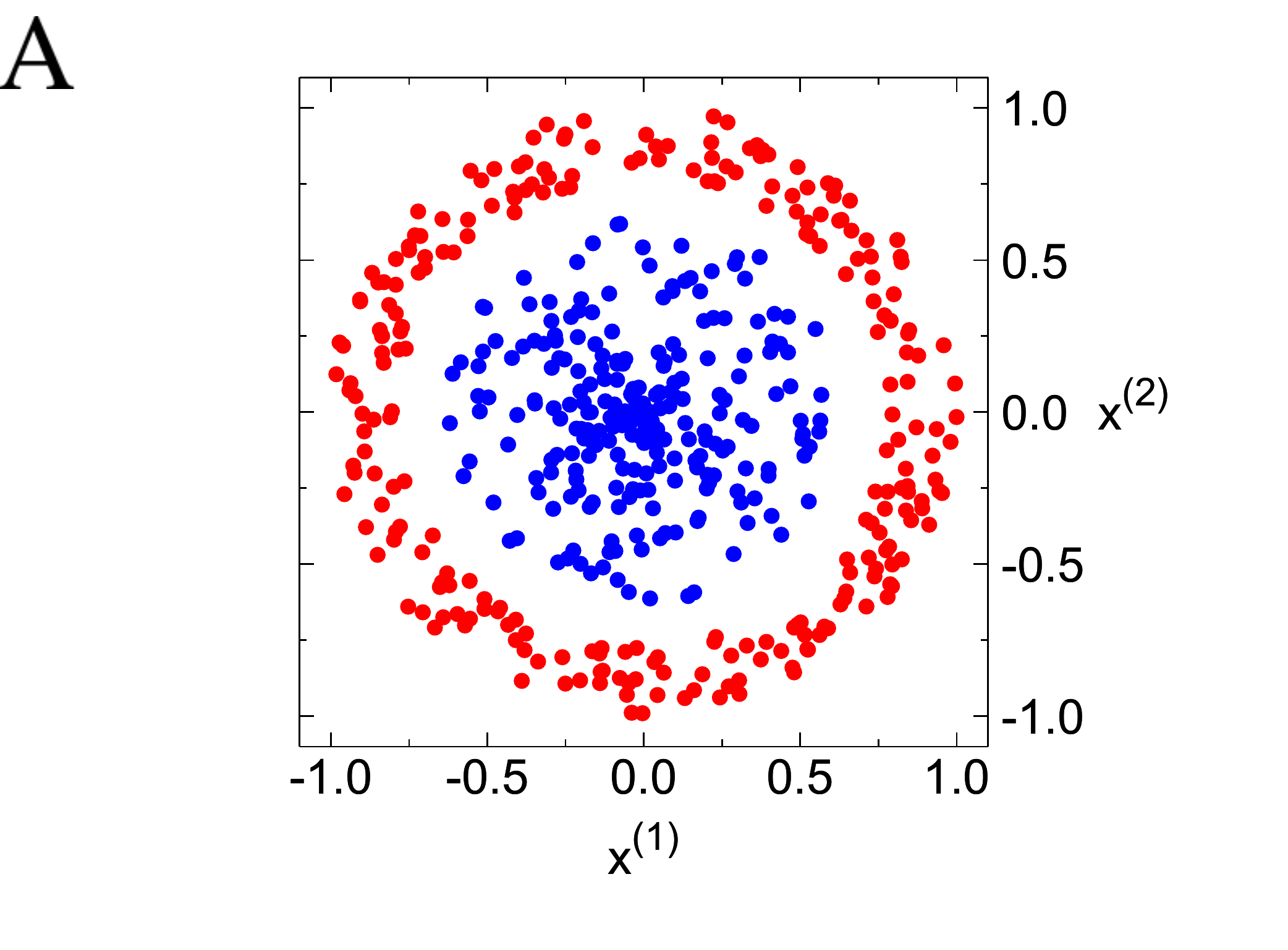}
\includegraphics[width=0.375\textwidth]{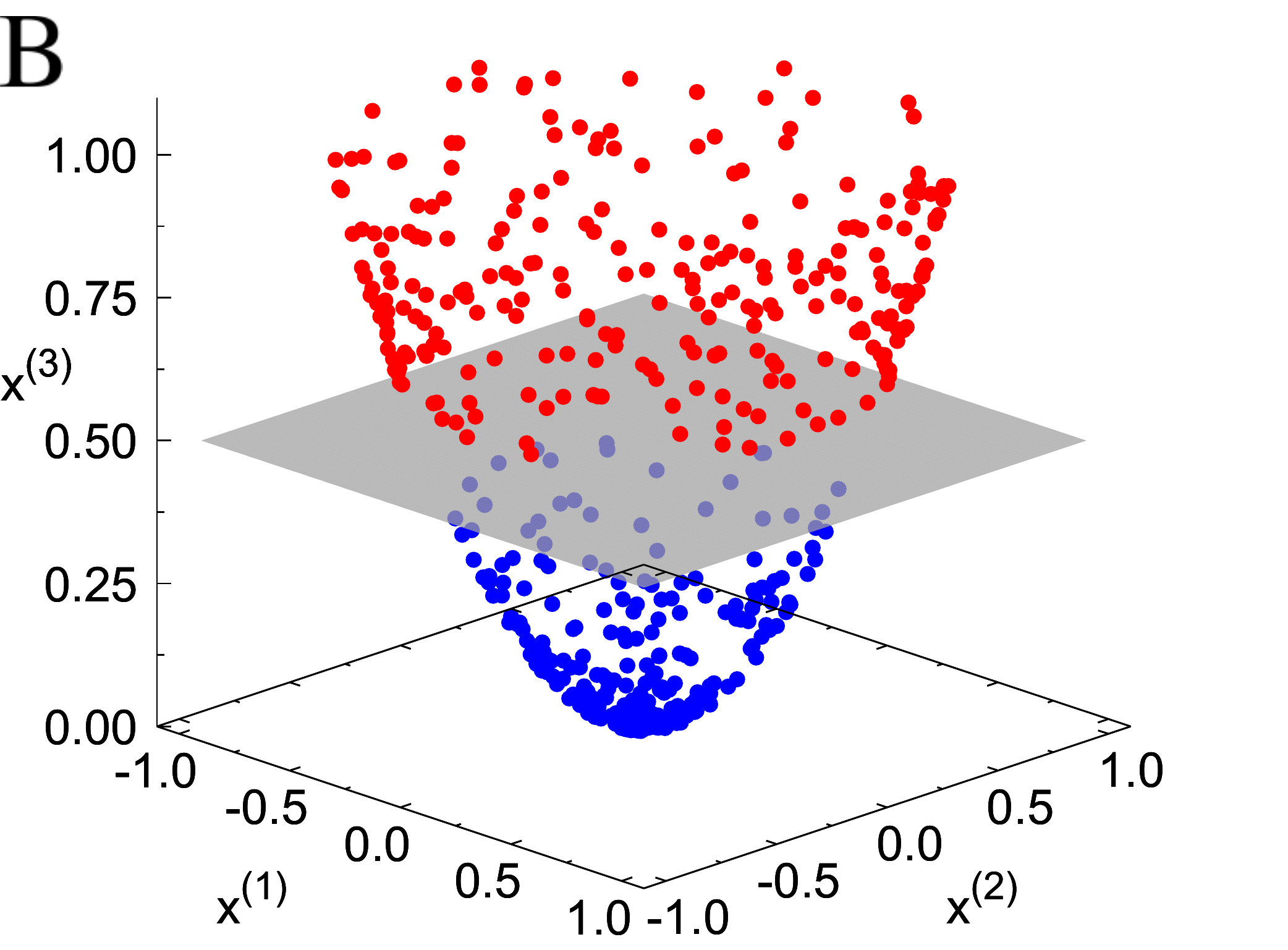}
\caption{\textbf{A}: The blue and red points with coordinates
  ($x^{(1)}$ , $x^{(2)}$) are linearly inseparable. \textbf{B}: By
  defining a suitable mapping from the input space
  ($x^{(1)}$,$x^{(2)}$) to a higher-dimensional feature space
  ($x^{(1)}$,$x^{(2)}$;,$x^{(3)}$), blue and red points become
  linearly separable by a plane at $x^{(3)} = 0.5$ (grey).}
\label{fig:linear_separability_in_feature_space}
\end{figure}

\noindent
Most ML methods used for PES construction are either kernel-based or
rely on artificial neural networks (ANNs). Both variants take
advantage of the fact that many nonlinear problems (such as predicting
energy from nuclear positions) can be linearised by mapping the inputs
to a (often higher-dimensional) feature space (see
Fig.~\ref{fig:linear_separability_in_feature_space}).\cite{scholkopf1997kernel}
Kernel-based methods utilize the \textit{kernel
  trick},\cite{boser1992training,scholkopf1998nonlinear,theodoridis2008pattern}
which allows to operate in an implicit feature space without
explicitly computing the coordinates of data in that space (see
section~\ref{sec:reproducing_kernel_representations} for more
details). ML methods based on ANNs rely on ``neuron layers'', which
map their inputs to feature spaces by linear transformations with
learnable parameters, followed by a nonlinearity (called activation
function). Often, many such layers are stacked on top of each other to
build increasingly complex feature spaces (see
section~\ref{sec:artificial_neural_networks}). In the following, both
variants are discussed in more detail.\\

\subsection{Reproducing Kernel Representations}
\label{sec:reproducing_kernel_representations}
Starting from a data set $\{(y_i; \mathbf{x}_i)\}_{i=1}^{N}$ of $N$
observations $y_i \in \mathbb{R}$ given the inputs $\mathbf{x}_i \in
\mathbb{R}^D$, kernel regression aims to estimate unknown values $y_*$
for inputs $\mathbf{x}_*$. For a PES, $y$ is typically the total
interaction energy and $\mathbf{x}$ is a representation of chemical
structure (i.e.\ a vector of internal coordinates, a molecular
descriptor like the Coulomb matrix\cite{rupp2012fast}, descriptors for
atomic environments, e.g. symmetry functions\cite{behler2011atom},
SOAP\cite{bartok2013representing} or
FCHL\cite{faber2018alchemical,christensen2019fchl}, or a
representation of crystal structure \cite{schutt2014represent,
  faber2015crystal,faber2016machine}). The representer
theorem\cite{scholkopf2001generalized} for a functional relation $y =
f(\mathbf{x})$ states that $f(\mathbf{x})$ can always be approximated
as a linear combination
\begin{equation}
f(\mathbf{x}) \approx \widetilde{f}(\mathbf{x}) = \sum_{i = 1}^{N} \alpha_i K(\mathbf{x},\mathbf{x}_i)
\label{eq:kernel_regression}
\end{equation}
where $\alpha_i$ are coefficients and $K(\mathbf{x},\mathbf{x'})$ is a
kernel function. A function $K(\mathbf{x},\mathbf{x'})$ is a
reproducing kernel of a Hilbert space $\mathcal{H}$ if the inner
product $\langle \phi(\mathbf{x}),\phi(\mathbf{x'})\rangle$ of
$\mathcal{H}$ can be expressed as
$K(\mathbf{x},\mathbf{x'})$.\cite{berlinet2011reproducing} Here,
$\phi$ is a mapping from the input space $\mathbb{R}^D$ to the Hilbert
space $\mathcal{H}$, i.e.\ $\phi: \mathbb{R}^D \mapsto
\mathcal{H}$. Many different kernel functions are possible, for
example the polynomial kernel
\begin{equation}
K(\mathbf{x},\mathbf{x'}) = \langle\mathbf{x},\mathbf{x'}\rangle^d
\label{eq:polynomial_kernel}
\end{equation}
where $\langle \cdot, \cdot \rangle$ denotes the dot product and $d$
is the degree of the polynomial, or the Gaussian kernel given by\\
\begin{equation}
K(\mathbf{x},\mathbf{x'}) = e^{-\gamma\lVert
  \mathbf{x}-\mathbf{x'}\rVert^2}
\label{eq:gaussian_kernel}
\end{equation}
are popular choices. Here, $\gamma$ is a hyperparameter determining
the width of the Gaussian and $\lVert \cdot \rVert$ denotes the
$L^2$-norm. It is also possible to include knowledge about the long
range behaviour of the physical interactions into the kernel function
itself\cite{hollebeek.annrevphychem.1999.rkhs} and the consequences of
such choices on the long- and short-range behaviour of the inter- and
extrapolation have been investigated in quite some
detail.\cite{hutson:2000}\\

\noindent
The mapping $\phi$ associated with the polynomial kernel
(Eq.~\ref{eq:polynomial_kernel}) depends on the dimensionality of the
inputs $\mathbf{x}$ and the chosen degree $d$ of the kernel. For
example, for $d=2$ and two-dimensional input vectors, the mapping is
given by $\phi:(x_1,x_2)\mapsto(x_1^2, \sqrt{2}x_1x_2, x_2^2)$ and the
Hilbert space $\mathcal{H}$ associated with the kernel function is
three-dimensional. For a Gaussian kernel, the associated Hilbert space
$\mathcal{H}$ even is $\infty$-dimensional. This can easily be seen if
Eq.~\ref{eq:gaussian_kernel} is rewritten as
\begin{equation}
K(\mathbf{x},\mathbf{x'}) = e^{-\gamma\lVert \mathbf{x}\rVert^2}e^{-\gamma\lVert \mathbf{x'}\rVert^2}e^{2\gamma \langle\mathbf{x},\mathbf{x'}\rangle}
\label{eq:gaussian_kernel_rewritten}
\end{equation}
then the Taylor expansion of the third factor
$e^{2\gamma\langle\mathbf{x},\mathbf{x'}\rangle}=\sum_{d=0}^\infty\frac{1}{d!}\left(
2\gamma\langle\mathbf{x},\mathbf{x'}\rangle\right)^d$ reveals that the
Gaussian kernel is equivalent to an infinite sum over polynomial
kernels (scaled by constant terms). It is important to point out that
in order to apply Eq.~\ref{eq:kernel_regression}, the mapping $\phi$
has never to be calculated explicitly (or even known at all) and it is
therefore possible to operate in the (high-dimensional) space
$\mathcal{H}$ implicitly. This is often referred to as the
\textit{kernel
  trick}\cite{boser1992training,scholkopf1998nonlinear,theodoridis2008pattern}.\\

\noindent
The coefficients $\alpha_i$ (Eq.~\ref{eq:kernel_regression}) can be
determined such that $\widetilde{f}(\mathbf{x}_i) = y_i$ for all
inputs $\mathbf{x}_i$ in the dataset, i.e.\
\begin{equation}
	\boldsymbol{\alpha} = \mathbf{K}^{-1}\mathbf{y}
	\label{eq:krr_coefficient_relation}
\end{equation}
where $\boldsymbol{\alpha} = \left[\alpha_i \cdots
  \alpha_N\right]^\mathrm{T}$ is the vector of coefficients,
$\mathbf{K}$ is an $N\times N$ matrix with entries $K_{ij} =
K(\mathbf{x}_i,\mathbf{x}_j)$ called kernel
matrix\cite{muller2001introduction,hofmann2008kernel} and $\mathbf{y}
= \left[y_1 \cdots y_N\right]^\mathrm{T}$ is a vector containing the
$N$ observations $y_i$ in the data set. Since the kernel matrix is
symmetric and positive-definite by construction, the efficient
Cholesky decomposition\cite{golub2012matrix} can be used to solve
Eq.~\ref{eq:krr_coefficient_relation}.  Once the coefficients
$\alpha_i$ have been determined, unknown values $y_*$ at arbitrary
positions $\mathbf{x}_*$ can be estimated as
$y_*=\widetilde{f}(\mathbf{x}_*)$ using
Eq.~\ref{eq:kernel_regression}. \\

\noindent
In practice however, the solution of
Eq.~\ref{eq:krr_coefficient_relation} is only possible if the kernel
matrix $\mathbf{K}$ is well-conditioned. Fortunately, in case
$\mathbf{K}$ is ill-conditioned, a regularized solution can be
obtained for example by Tikhonov
regularization\cite{tikhonov1977solutions}. This amounts to adding a
small positive constant $\lambda$ to the diagonal of $\mathbf{K}$,
such that
\begin{equation}
    \boldsymbol{\alpha} = \left(\mathbf{K}+\lambda\mathbf{I}\right)^{-1}\mathbf{y}
	\label{eq:krr_coefficient_relation_regularized}
\end{equation}
is solved instead of Eq. \ref{eq:krr_coefficient_relation} when
determining the coefficients $\alpha_i$ (here, $\mathbf{I}$ is the
identity matrix). For non-zero $\lambda$ however,
$\widetilde{f}(\mathbf{x}_i) \neq y_i$ and
Eq.~\ref{eq:kernel_regression} reproduces the known values in the data
set only approximately. Therefore, this method of determining the
coefficients can also be used to prevent over-fitting and is known as
kernel ridge regression (KRR).\cite{rupp2015machine}\\

\noindent
KRR is closely related to Gaussian process regression
(GPR).\cite{gp:2005} In GPR, it is assumed that the $N$ observations
$\{(y_i; \mathbf{x}_i)\}_{i=1}^{N}$ in the data set are generated by a
Gaussian process, i.e.\ drawn from a multivariate Gaussian
distribution with zero mean, and covariance
$K(\mathbf{x},\mathbf{x'})$. Note that a mean of zero can always be
assumed without loss of generality since two multivariate Gaussian
distributions with equal covariance matrix can always be transformed
into each other by addition of a constant term. Further, every
observation $y_i$ is considered to be related to $\mathbf{x}_i$
through an underlying function $f(\mathbf{x})$ and some observational
noise (e.g.\ due to uncertainties in measuring $y_i$)
\begin{equation}
y_i =  f(\mathbf{x}_i) + \mathcal{N}(0,\lambda)
\label{eq:gaussian_process_noise_model}
\end{equation}
where $\lambda$ is the variance of the Gaussian noise model. The
chosen covariance function $K(\mathbf{x},\mathbf{x'})$ expresses an
assumption about the nature of $f(\mathbf{x})$. For example, if the
Gaussian kernel (Eq.~\ref{eq:gaussian_kernel}) is used,
$f(\mathbf{x})$ is assumed to be smooth and the chosen Gaussian width
$\gamma$ determines how rapid $f(\mathbf{x})$ is allowed to change if
the input $\mathbf{x}$ changes.\\

\noindent
With these assumptions, it is now possible to determine the
conditional probability $p(y_*|\mathbf{y})$, i.e.\ answer the question
``given the data $\mathbf{y}=\left[y_1 \cdots y_N\right]^\mathrm{T}$,
how likely is it to observe the value $y_*$ for an input
$\mathbf{x}_*$?''. Since it was assumed that the data was drawn from a
multivariate Gaussian distribution, it is possible to write
\begin{equation}
	\begin{bmatrix}
	\mathbf{y}\textbf{}\\
	y^*
	\end{bmatrix}
	\sim
	\mathcal{N}\left(0,
	\begin{bmatrix}
    \mathbf{K} + \lambda\mathbf{I} & \mathbf{K}_*^\mathrm{T}\\
    \mathbf{K}_* &  K(\mathbf{x}_*,\mathbf{x}_*)
	\end{bmatrix}
	\right)
\label{eq:gaussian_process_distribution}
\end{equation}
where $\mathbf{K}$ is the kernel matrix (see
Eq.~\ref{eq:krr_coefficient_relation}) and $\mathbf{K}_* =
\left[K(\mathbf{x}_*,\mathbf{x}_1) \cdots
  K(\mathbf{x}_*,\mathbf{x}_N)\right]$.  Then, the best (most likely)
estimate for $y_*$ is the mean of this distribution
\begin{equation}
\Bar{y}_* = \mathbf{K}_* \left(\mathbf{K} + \lambda\mathbf{I}\right)^{-1}\mathbf{y}
\label{eq:gaussian_process_mean}
\end{equation}
Thus, estimating $y_*$ with GPR (Eq.~\ref{eq:gaussian_process_mean})
is mathematically equivalent to estimating $y_*$ with KRR (compare to
Eqs.~\ref{eq:kernel_regression}~and~\ref{eq:krr_coefficient_relation_regularized}). However,
while in KRR, $\lambda$ is only a hyperparameter related to
regularization, in GPR, $\lambda$ is directly related to the magnitude
of the assumed observational noise (see
Eq.~\ref{eq:gaussian_process_noise_model}). Further, the predictive
variance,
\begin{equation}
\mathrm{var}(y_*) = K(\mathbf{x}_*,\mathbf{x}_*) - \mathbf{K}_* \left(\mathbf{K} + \lambda\mathbf{I}\right)^{-1} \mathbf{K}_*^\mathrm{T}
\label{eq:gaussian_process_variance}
\end{equation}
which can also be derived from
Eq.~\ref{eq:gaussian_process_distribution}, can be useful to estimate
the uncertainty of a prediction $y_*$, i.e.\ how confident the model
is that its prediction is correct. Since KRR and GPR are so similar,
they are both referred to as reproducing kernel representations in
this work.\\

\subsection{Artificial Neural Networks}
\label{sec:artificial_neural_networks}
The basic building blocks of artificial neural networks
(NNs)\cite{mcculloch1943logical,kohonen1988introduction,abdi1994neural,bishop1995neural,clark1999neural,ripley2007pattern,haykin2009neural}
are so-called ``dense (neuron) layers'', which transform input vectors
$\mathbf{x}\in \mathbb{R}^{n_{\rm in}}$ linearly to output vectors
$\mathbf{y}\in \mathbb{R}^{n_{\rm out}}$ through
\begin{equation}
\mathbf{y} = \mathbf{W}\mathbf{x} + \mathbf{b}
\label{eq:dense_layer}
\end{equation}
where the weights $\mathbf{W}\in \mathbb{R}^{n_{\rm out} \times n_{\rm
    in}}$ and biases $\mathbf{b}\in \mathbb{R}^{n_{\rm out}}$ are
parameters, and $n_{\rm in}$ and $n_{\rm out}$ denote the
dimensionality of inputs and outputs, respectively. A single dense
layer can therefore only represent linear relations. To model
non-linear relationships between inputs and outputs, at least two
dense layers need to be combined with a non-linear function $\sigma$
(called activation function), i.e.
\begin{eqnarray}
\label{eq:single_layer_hidden}
\mathbf{h} =& \sigma\left(\mathbf{W}_1\mathbf{x} + \mathbf{b}_1\right)\\
\label{eq:single_layer_output}
\mathbf{y} =& \mathbf{W}_2\mathbf{h} + \mathbf{b}_2
\end{eqnarray}
Such an arrangement
(Eqs.~\ref{eq:single_layer_hidden}~and~\ref{eq:single_layer_output})
has been proven to be a general function approximator, meaning that
any mapping between input $\mathbf{x}$ and output $\mathbf{y}$ can
be approximated to arbitrary precision, provided that the
dimensionality of the so-called ``hidden layer'' $\mathbf{h}$ is large
enough.\cite{gybenko1989approximation,hornik1991approximation} As
such, NNs are a natural choice for representing a PES, i.e.\ a mapping
from chemical structure to energy (for PES construction, the output
$\mathbf{y}$ usually is one-dimensional and represents the energy).\\

\noindent
While \textit{shallow} NNs with a single hidden layer (see above) are
in principle sufficient to solve any learning task, in practice,
\textit{deep} NNs with multiple hidden layers are exponentially more
parameter-efficient.\cite{eldan2016power} In a deep NN, $l$ hidden
layers are stacked on top of each other,
\begin{equation}
\begin{aligned}
\mathbf{h}_1 &= \sigma\left(\mathbf{W}_1\mathbf{x} + \mathbf{b}_1\right)\\
\mathbf{h}_2 &= \sigma\left(\mathbf{W}_2\mathbf{h}_1 + \mathbf{b}_2\right)\\
&\vdots\\
\mathbf{h}_l &= \sigma\left(\mathbf{W}_l\mathbf{h}_{l-1} + \mathbf{b}_l\right)\\
\mathbf{y} &= \mathbf{W}_{l+1}\mathbf{h}_l + \mathbf{b}_{l+1}
\end{aligned}
\end{equation}
mapping the inputs $\mathbf{x}$ to increasingly complex feature
spaces, until the features $\mathbf{h}_l$ in the final layer are linearly related to the outputs $\mathbf{y}$. The parameters of the NN, i.e.\ the entries in the matrices $\mathbf{W}_l$ and vectors $\mathbf{b}_l$, are initialized randomly and then optimized, for
example via gradient descent, to minimize a loss function that
measures the difference between the output of the NN and a given set
of training data. For example, the mean squared error (MSE) is a
popular loss function for regression tasks.\\

\noindent
The earliest NN-based PESs directly use a set of internal
coordinates, e.g.\ distances and angles, as input for the
NN.\cite{blank1995neural,brown1996combining,tafeit1996neural,no1997description,prudente1998fitting}
However, such approaches have the disadvantage that swapping symmetry
equivalent atoms may also change the numeric values of the internal
coordinates. Since it is not guaranteed that a NN maps two different
inputs related by a permutation operation to the same output energy,
the permutational invariance of the PES is violated. Another
disadvantage of using internal coordinates as input is that a NN
trained for a single molecule cannot be used to calculate the energy
of a dimer, because they require a different number of internal
coordinates for an unambiguous description of the molecular
geometry. Therefore, for small systems, PESs based on NNs have been
designed in the spirit of a many-body
expansion,\cite{manzhos2006random,manzhos2007using,malshe2009development}
which circumvents these issues. However, these approaches involve a
large number of individual NNs, i.e.\ one for each term in the
many-body expansion and scale poorly for large systems.\\

\noindent
For larger systems, it is common practice to decompose the total
energy of a chemical system into atomic contributions, which are
predicted by a single NN (or one for each element). This approach,
known as high-dimensional neural network
(HDNN)\cite{behler2007generalized} and first proposed by Behler and
Parrinello, relies on the chemically intuitive assumption that the
contribution of an atom to the total energy depends mainly on its
local environment.\\

\noindent
Two variants of HDNNs can be distinguished: The ``descriptor-based''
variant uses a hand-crafted
descriptor,\cite{behler2011atom,khorshidi2016amp,artrith2017efficient,unke2018reactive}
to encode the environment of an atom, which is then used as input of a
multi-layer feed-forward NN. Examples for this kind of approach are
ANI\cite{smith2017ani} and TensorMol.\cite{yao2018tensormol} The
``message-passing''\cite{gilmer2017neural} variant directly uses
nuclear charges and Cartesian coordinates as input and a deep neural
network (DNN) is used to exchange information (``messages'') between
individual atoms, such that a
representation of their chemical environments is learned directly from
the data. The DTNN\cite{schutt2017quantum} introduced by Sch\"utt
\textit{et al.}  was the first NN of this kind and has since been
refined in other DNN architectures, for example
SchNet\cite{schutt2017schnet}, HIP-NN\cite{lubbers2018hierarchical} or
PhysNet\cite{unke2019physnet}. Both types of HDNN perform well,
however, the message-passing variant is able to automatically adapt
the description of the chemical environments to the training data and
the prediction task at hand and usually achieves a better
performance.\cite{schutt2018quantum}

\section{Applications}
In the following, illustrative applications of explicit
representations of PESs (see
section~\ref{sec:explicit_representations}) and machine-learned PESs
(see section~\ref{sec:machine_learned_pess}) are discussed. Potential
energy surfaces of sufficient quality for gas- and solution-phase
reactions differ in at least two respects. While for reactions in the
gas phase, typically involving small molecules as reactants,
techniques to construct global, reactive PESs are becoming available,
this is not so for reactions in solutions. Often, the global property
is also not required {\it a priori} for reactions in
solution. Secondly, for reactions in the gas phase all interactions
are typically encoded in the global, reactive PES itself, whereas for
reactions in solution the interaction between solute and solvent needs
to be represented separately and explicitly. Therefore gas- and
solution-phase are discussed in two different sections
~\ref{sec:gas_phase_reaction_dynamics}~and~\ref{sec:reactions_in_the_condensed_phase}. While
PESs are often used to explore the conformational space of a given
system or study molecular (reaction) dynamics, machine-learned PESs
can also serve as an alternative to \textit{ab initio} methods for
exploring chemical compound space. For example, it is possible to
predict energies of molecules of different chemical composition from
learning on a reference data set. Such applications are discussed
briefly in section~\ref{sec:energy_predictions}.\\

\subsection{Gas Phase Reaction Dynamics}
\label{sec:gas_phase_reaction_dynamics}
\paragraph{Multisurface, reactive dynamics for triatomics:} Triatomic
systems constitute an important class of systems relevant to the
chemistry in the hypersonic regime. Typical reactive collisions upon
reentry of objects from outer space into Earth's atmosphere include
the O+NO, O+CO, N+CO, C+NO, or N+NO systems. Due to the high
velocities of the impacting object, temperatures up to 20000~K can be
reached. To study the reaction dynamics at such high collision
energies both, ground and lower electronically excited states need to
be included. Hence, to describe the reactive dynamics for such
systems, fully dimensional, reactive PESs including multiple
electronic states are required. This is possible by using a large
number of {\it ab initio} calculated energies at the multi-reference
CI level of theory and representing the PESs using a reproducing
kernel Hilbert space (RKHS). Alternative approaches use explicit
fitting of a parametrized form of a suitable many body expansion of
the PES.\cite{dmbe:1988}\\

\begin{figure}
\centering
\includegraphics[scale=1.3]{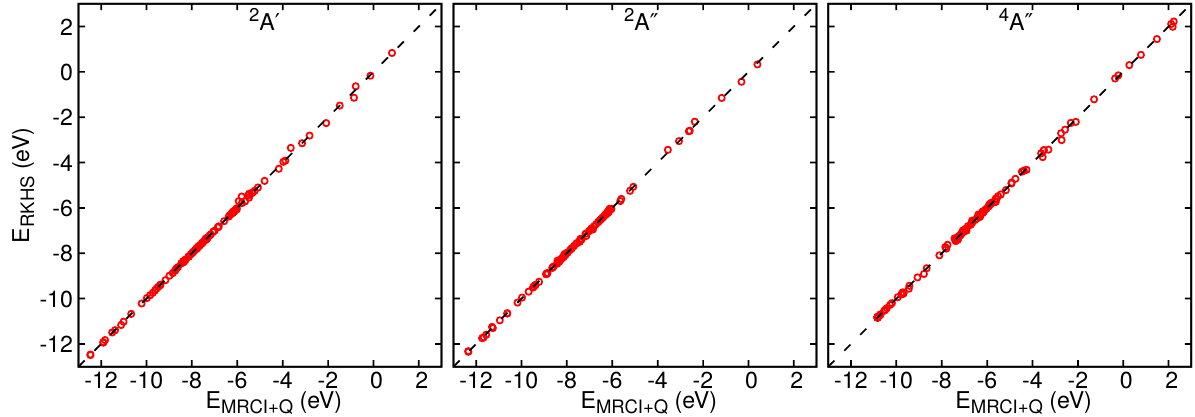}
\caption{Correlation between the RKHS and MRCI+Q energies for test
  data set for $^2$A$'$, $^2$A$''$ and $^4$A$''$ electronic state of
  CNO system. Black dashed line shows the diagonal. Data taken form
  Ref. \cite{MM.cno:2018}}
\label{fig:cnocorr}
\end{figure}

\begin{figure}
\centering \includegraphics[scale=0.65]{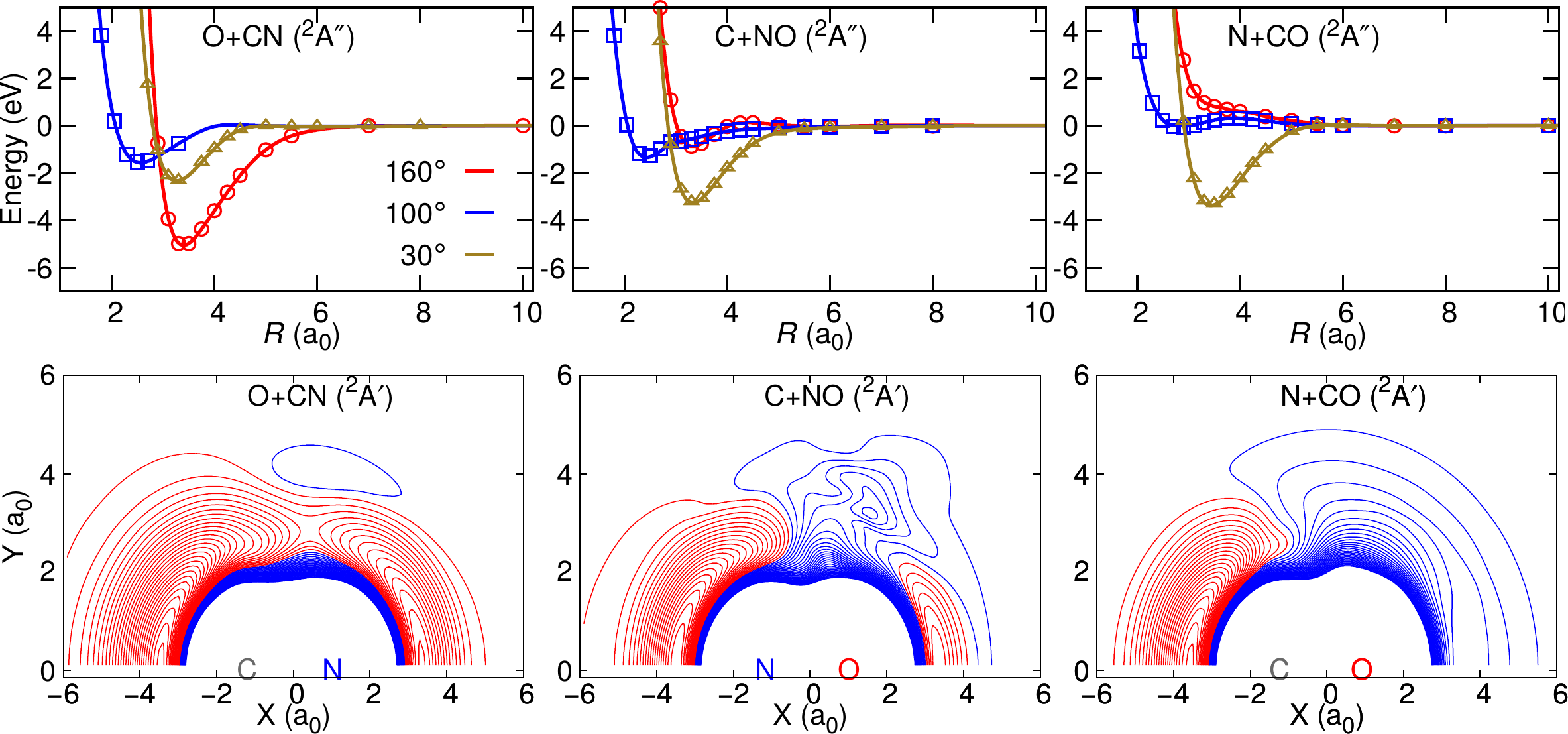}
\caption{Upper panel: Comparison between the RKHS (solid lines) and
  MRCI+Q (open symbols) energies for different 1D cuts in Jacobi
  coordinates in the O+CN, C+NO and N+CO channels for the $^2$A$''$
  electronic state of CNO system. Lower panel: Contour diagram of the
  3D RKHS PESs for three different channels of CNO $^2$A$'$
  system. The diatoms are fixed to their equilibrium geometry and the
  zero is set to the asymptotic value for each channels. Data taken
  form Ref. \cite{MM.cno:2018}}
\label{fig:cnocomp}
\end{figure}

\noindent
One example for such a system constitutes the reactive dynamics of
[CNO] in the hypersonic regime at temperatures up to $T = 20000$
K.\cite{MM.cno:2018} The C+NO reaction is important in combustion
chemistry and NO plays a crucial role in the chemistry near the
surface of a space vehicle during atmospheric re-entry.\cite{per15:30}
For this, accurate fully dimensional PESs for the $^2$A$'$, $^2$A$''$
and $^4$A$''$ states were determined and used in quantum dynamics and
quasiclassical trajectory simulations. More than 50000 {\it ab initio}
energies were calculated at the MRCI+Q/aug-cc-pVTZ level of theory to
construct the RKHS PESs. The electronic structure calculations were
performed in grids based on Jacobi coordinates for each channels. RKHS
was used to construct analytical representations for each channel and
global 3D PES was then made by smoothly connecting the PESs for the
three channels by switching functions. Correlation plots of the MRCI+Q
energies and the analytical energies obtained from the 3D RKHS based
PESs are shown in Figure \ref{fig:cnocorr} for three sets of off-grid
points calculated to validate the quality of the RKHS-based global
PESs. RKHS energies for different 1D cuts are compared with {\it ab
  initio} energies in all three channels for the $^2$A$''$ PES in
Figure \ref{fig:cnocomp}. The contour plots shown in Figure
\ref{fig:cnocomp} shows the topology of the $^2$A$'$ PES for all three
channels. The overall good agreement between the {\it ab initio} and
analytical energies in all the channels and for all the electronic
states suggests the high quality of the PESs.\\

\begin{figure}
\centering
\includegraphics[scale=1.00]{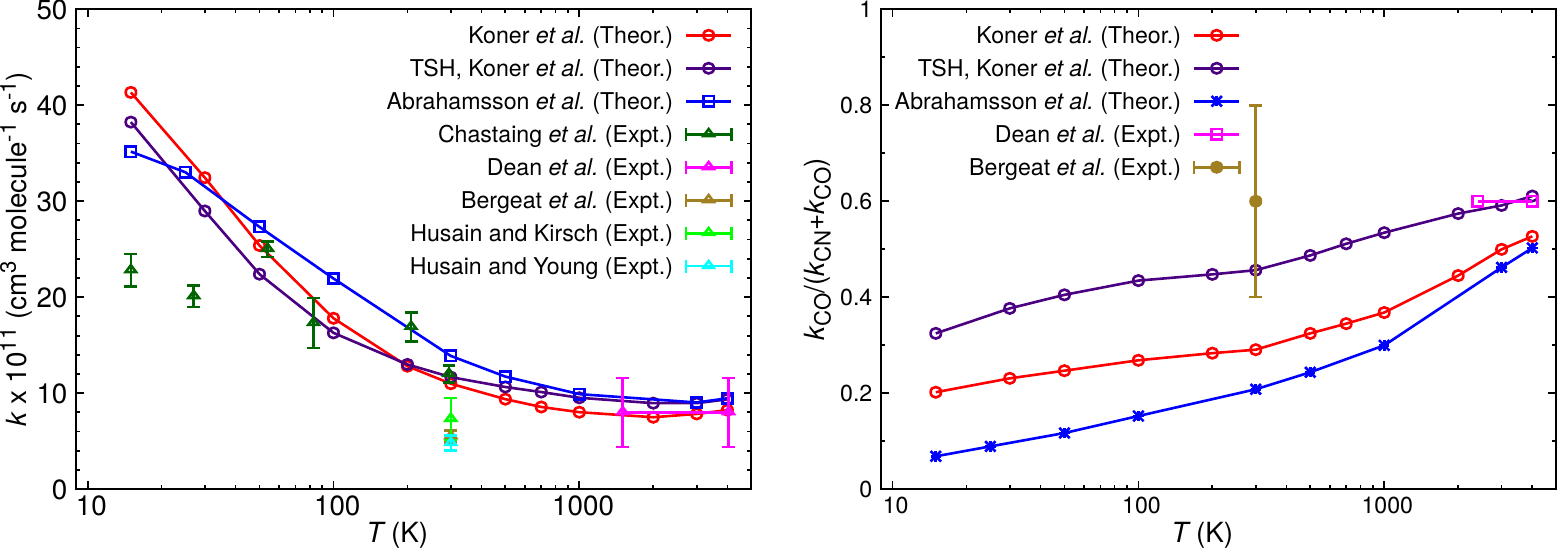}
\caption{Total rate coefficients (left) and branching fractions
  (right) C($^3$P) + NO(X$^2\Pi$) $\rightarrow$ O($^3$P) +
  CN(X$^2\Sigma^+$) and N($^2$D)/N($^4$S) + CO(X$^1\Sigma^+$) reaction
  compared with available experimental and theoretical data. Data
  taken form Ref. \cite{MM.cno:2018}}
\label{fig:cnorates}
\end{figure}

\noindent
Experimental reference data is available for the rate coefficients and
branching fractions of CO and CN products for the the C($^3$P) +
NO(X$^2\Pi$) $\rightarrow$ O($^3$P) + CN(X$^2\Sigma^+$) and
N($^2$D)/N($^4$S) + CO(X$^1\Sigma^+$)
reaction.\cite{dea91:3180,ber99:7,bra69:417} From 40000 QCT
trajectories at each temperature on each PES run both, in an adiabatic
and a nonadiabatic fashion within a
Landau-Zener\cite{lan32:46,zen32:696,bel11:014701,bel14:224108}
formalism, the rate coefficients and branching fractions were
determined. These rates can be directly compared with experiments and
previous simulations. Figure \ref{fig:cnorates} shows the rate
coefficients and branching ratios for the products. Except for the
lowest temperatures ($T \sim 30$ K and below) the rate coefficients
agree well with experiments. Furthermore, it was found that including
nonadiabatic transitions leads to better agreement with experiment
within error bars but without nonadiabatic transitions the branching
fractions were underestimated, see right panel in Figure
\ref{fig:cnorates}. In addition, computed final state distributions of
the products for molecular beam-type simulations agree well with
experiment. From such studies, thermal rates within an Arrhenius
formalism can be determined which can then be used in more coarse
grained simulations, such as discrete sampling Monte Carlo
(DSMC).\cite{dsmc}\\

\noindent
\paragraph{Reactive dynamics of larger gas-phase systems:} One recent
application of MS-ARMD and a NN-trained PES concerned the Diels-Alder
reaction between 2,3-dibromo-1,3-butadiene (DBB) and maleic anhydride
(MA).\cite{MM.diels:2019} DBB is a generic diene which fulfills the
experimental requirements for conformational separation of its isomers
by electrostatic deflection of a molecular
beam,\cite{chang13a,willitsch17a} thus enabling the characterization
of conformational aspects and specificities of the reaction. MA is a
widely used, activated dienophile which due to its symmetry simplifies
the possible products of the reaction. The reaction of DBB and MA thus
serves as a prototypical system well suited for the exploration of
general mechanistic aspects of Diels-Alder processes in the gas
phase. The main questions concerned the synchronicity and
concertedness of the reaction and how the reaction could be
promoted. Until now, computational studies of Diels-Alder reactions
including the molecular dynamics have started from TS-like
structures\cite{souza16a,black12a,tan18a,wang09a,liu16a} or have used
steered dynamics\cite{soto16a} both of which introduce biases and do
not allow direct calculation of reaction rates.\\

\noindent
To do so, two different reactive PESs were developed. One was based on
the MS-ARMD approach whereas the second one employed the
PhysNet\cite{unke2019physnet} architecture to train a NN
representation. For both representations scattering calculations were
started from suitable initial conditions by sampling the internal
degrees of freedom of the reactants and the collision parameter
$b$. It is found that the majority of reactive collisions occur with
rotational excitation and that most of them are synchronous. The
relevance of rotational degrees of freedom to promote the reaction was
also found when the minimum dynamical path\cite{unke2019sampling} was
calculated. The dynamics on both, the MS-ARMD and NN-trained PESs are
very similar although the quality of the two surfaces is
different. While the NN-trained PES is able to reproduce the training
data to within 0.25 kcal/mol on average, the RMSD between reference
and parametrized PES for MS-ARMD is 1.5 kcal/mol over a range of 80
kcal/mol. In terms of computational efficiency MS-ARMD is about 200
times faster than PhysNet.\\

\noindent
Another prototypical example of a chemical reaction is the
S$_\mathrm{N}$2 mechanism. In a recent comparative
study,\cite{brickel2019reactive} three reactive PESs for the
[Cl--CH$_3$--Br]$^-$ system were constructed: Two of these PESs rely
on empirical force fields, either combined with the MS-ARMD or the
MS-VALBOND\cite{mm.msvalb:2018} approach to construct the global
reactive PES, whereas the third is NN-based. While all methods are
able to fit the \textit{ab initio} reference data with $R^2 > 0.99$,
the NN-based PES achieves mean absolute and root mean squared
deviations that are an order of magnitude lower than the other methods
when using the same number of reference data. When increasing the size
of the reference data set, the prediction errors made by the NN-based
PES are even up to three orders of magnitude lower than for the force
field-based PESs. However, at the same time, evaluating the NN-based
PES is about three orders of magnitude
slower.\cite{brickel2019reactive}\\

\subsection{Reactions in the Condensed Phase}
\label{sec:reactions_in_the_condensed_phase}
For reactions in the condensed phase, two different situations are
considered in the following. In one of them, ligands bind to a
substrate anchored within a protein, such as for small diatomic
ligands binding to the heme-group in globins. In the other, the
substrate is chemically transformed as is the case for the Claisen
rearrangement from chorismate to prephenate.\\

\noindent
\paragraph{Ligand (Re-)Binding in Globins:} Computationally, the structural
dynamics accompanying NO-rebinding to Myoglobin has recently been
investigated with the aim to assign the transient, metastable
structures relevant for rebinding of the ligand on different time
scales.\cite{meuwly.2016.mbno} For this, reactive MD simulations using
MS-ARMD simulations were run involving the bound $^2A$ and the unbound
$^4A$ states which are also probed experimentally. The energy for each
of the states was represented as a reproducing
kernel\cite{meuwly.2016.mbno,hollebeek.annrevphychem.1999.rkhs,MM.rkhs:2017}
for the subspace of important system coordinates (the heme(Fe)--NO
separation and angle, and the doming coordinate of the heme-Fe)
combined with an empirical force field for all remaining degrees of
freedom. Such an approach is inspired by a decomposition of the system
into a region that is modelled with high accuracy (typically a
``quantum region'') and an environment (the ``molecular mechanics''
part).\\

\noindent
With a system parametrized in this fashion, extensive reactive MD
simulations were run.\cite{meuwly.2016.mbno} The kinetics for ligand
rebinding is nonexponential with time scales of 10 and 100 ps. These
are consistent with time scales measured from optical, infrared, and
X-ray absorption experiments and previous computational
work.\cite{cornelius83mbno,Petrich1991,Ionascu05p16921,Kruglik2010,lim2012mbno,negrerie2012mbno,silatani:2015,Kim2004,Meuwly02p183,champion2002,Nutt06p1191,danielsson-jctc-08}
The influence of the iron-out-of-plane (Fe-oop or ``doming'')
coordinate on the rebinding reaction, as predicted by
experiment\cite{Kruglik2010}, was directly established. The two time
scales (10 and 100 ps) are associated with two structurally different
states of the His64 side chain -- one ``out'' (state A) and one ``in''
(state B) -- which control ligand access and rebinding dynamics. Such
an unequivocal assignment was not possible from
experiment.\cite{Merchant03} In addition, the simulations provide an
explanation why an energetically feasible state for NO-binding to heme
is typically not found in Mb: Although the bound Fe-ON state is a
local minimum on the potential energy surface, the energy of this
state on the unbound $^4A$ manifold is lower and, hence, the bound
$^2A$ Fe-ON can not be spectroscopically characterized. The
simulations finally clarify that the XAS experiments are unable to
distinguish between structures with photodissociated NO ``close to''
or ``far away'' from the heme-Fe in the active site as had been
proposed.\cite{silatani:2015}\\

\noindent
In this fashion, validation of experimental results by the MD
simulations and in-depth analysis of the configurations driving the
dynamics on the different time scales (10 ps and 100 ps) allowed to
identify the structural origins of the conformational dynamics at a
molecular level. It is expected that further combined experimental and
computational studies of this kind will provide the necessary insight
to link energetics, structures and dynamics in complex systems.\\

\noindent
\paragraph{Reactions in Solution:} The Claisen
rearrangement\cite{claisen.ber.1912.general} is an important
[3,3]-sigmatropic rearrangement for high
stereoselective\cite{iwakura.pccp.2011.general} C-C bond
formation\cite{coates.jacs.1987.ts}. The text book example of a
Claisen rearrangement is the reaction of allyl-vinyl ether (AVE) to
pent-4-enal\cite{ziegler.chemrev.1988.general}. In polar solvent the
stabilization of the transition state (TS) relative to the reaction in
vacuum is the origin of the catalytic
effect.\cite{severance.jacs.1992.ave,guest.perk2.1997.ave,Cramer1992}
This has motivated numerous studies on enzymatic Claisen
rearrangements in
particular\cite{kast.tetl.1996.chorismate,Ranaghan2003,Lever2014,Marti2010,Ferrer2011,Marti2001,Roca2009,madurga.pccp.2001.chorismate,davidson.perk2.1997.chorismate,wiest.jacs.1995.chorismate,Hur2002}
and reactions with related
substrates\cite{vance.jacs.1988.ts,claeyssens.orgbiochem.2011.chorismate,ranaghan.orgbiochem.2004.chorismate,carlson.jacs.1996.chorismate}. Compared
to the reaction in aqueous solution the enzymatic catalysis of the
Claisen rearrangement reaction in chorismate mutase (CM) leads to a
rate acceleration by $\sim 10^6$ due to stabilisation of the
TS.\cite{Andrews1973}\\

\noindent
A reactive force field based on MS-ARMD was parametrized for AVE and
used unchanged for AVE-(CO$_2$)$_2$ and chorismate to study their
Claisen rearrangements in the gas phase, in water and in the
chorismate mutase from \textit{Bacillus subtilis} (BsCM). Using free
energy simulations it is found that in going from AVE and
AVE-(CO$_2$)$_2$ to chorismate and using the same reactive PES the
rate slows down when going from water to the protein as the
environment. However, for the largest substrate (chorismate) they
correctly find that the protein accelerates the reaction. Considering
the changes of $+4.6$ (AVE), $+2.9$ (AVE-(CO$_2$)$_2$) and $-4.4$
(chorismate) kcal/mol in the activation free energies and correlating
them with the actual chemical modifications suggests that both,
electrostatic stabilization (AVE$\rightarrow$AVE-(CO$_2)_2$) and
entropic contributions (AVE-(CO$_2$)$_2 \rightarrow$ chorismate,
through the rigidification and larger size of chorismate) lead to the
rate enhancement observed for chorismate in CM.\\

\noindent
As for the reaction itself it is found that for all substrates
considered the O-C bond breaks prior to C-C bond formation. This
agrees with kinetic isotope experiments according to which C-O
cleavage always precedes C-C bond formation.\cite{hilvert:2005} For
the nonenzymatic thermal rearrangement of chorismate to prephenate the
measured kinetic isotope effects\cite{Addadi1983,hilvert:2005}
indicate that at the TS the C-O bond is about 40 \% broken but little
or no C-C bond is formed, consistent with an analysis based on ``More
O'Ferrall-Jencks'' (MOFJ) diagrams.\cite{Ferrall1970,Jencks1972}\\

\noindent
The analysis of the TS position in the active site of BsCM reveals
that the lack of catalytic effect on AVE is due to its loose
positioning, insufficient interaction with and TS stabilization by the
active site of the enzyme. Major contributions to localizing the
substrate in the active site of BsCM originate from the CO$_2^-$
groups. This together with the probability distributions in the
reactant, TS and product states suggest that entropic factors must
also be considered when interpreting differences between the systems,
specifically (but not only) in the protein environment.\\

\subsection{Energy predictions}
\label{sec:energy_predictions}
The systematic exploration of chemical space is a possible way to find
as of yet unknown compounds with useful properties, e.g.\ for medical
applications. For example, the GDB-17
database\cite{ruddigkeit2012enumeration} enumerates 166 billion small
organic molecules that are potential drug candidates. However, running
\textit{ab initio} calculations to determine the properties of
billions of molecules is computationally infeasible. Machine-learned
PESs were shown to reach accuracies on par with hybrid DFT
methods\cite{faber2017prediction} and thus can serve as an efficient
alternative to predict e.g.\ stabilization energy or equilibrium
structures.\\

\noindent
In order to be able to compare different approaches, benchmark
datasets are used to assess the accuracy of ML methods. One of the
most popular benchmarks for this purpose is
QM9\cite{ramakrishnan2014quantum}. It consists of several properties
for 133'885 equilibrium molecules corresponding to the subset of all
species with up to nine heavy atoms (C, O, N, and F) out of the GDB-17
database calculated at the B3LYP/6-31G(2df,p) level of
theory.\cite{ruddigkeit2012enumeration} For example, after training on
50'000 structures, both the PhysNet neural network
architecture\cite{unke2019physnet} and KRR based on the FCHL2019
descriptor\cite{christensen2019fchl} achieve a mean absolute error of
$\approx0.3$~kcal/mol for predicting the energy of unseen
molecules. When the FCHL2018\cite{faber2018alchemical} descriptor is
used in the kernel model, the same accuracy is reached after training
on just 20'000 structures. However, FCHL2018 descriptors are
computationally expensive and therefore difficult to apply to larger
training set sizes.\cite{christensen2019fchl}\\

\noindent
It is also possible to predict other molecular properties (apart from
energy) with ML methods. Interested readers are referred to
Ref.~\citenum{faber2017prediction}, which compares the accuracy of
different approaches for the prediction of other properties, for
example HOMO/LUMO energies, dipole moments, polarizabilities, zero
point vibrational energies, or heat capacities. Since all molecular
properties can be derived from the wave function, recent approaches
aim to directly predict the electronic wave function from nuclear
coordinates\cite{schutt2019unifying} or incorporate response operators
into the model.\cite{christensen2019operators}\\

\section{Outlook and Conclusions}
This section puts the methods discussed in the present overview into
perspective and discusses future extensions, and their advantages and
disadvantages. \\

\noindent
As discussed, RKHS has been applied to generate accurate
representations of PES for different triatomic systems (3D) to study
either reactive collisions or vibrational spectroscopy. The RKHS
procedure can also be applied to construct higher dimensional PESs. As
an example, an RKHS representation of the 6D PES for N$_4$ is
discussed. Previously, a global PES was constructed for N$_4$ using
PIPs from 16435 CASPT2/maug-cc-pVTZ
energies\cite{TruhlarN4:2013,TruhlarN4Erratum:2014} which are also
used here. For constructing the RKHS, a total of 16046 {\it ab initio}
energies up to 1200 kcal/mol were used. The full PES is expanded in a
many body expansion,
\begin{equation}
    V(r_{i}) = \sum_{i=1}^4 V^{(1)}_i +  \sum_{i=1}^6 V^{(2)}_i(r_i) + \sum_{i=1}^3 V^{(3)}_i(r_j,r_k,r_l) + V^{(4)}_i(r_j,r_k,r_l,r_m)
\end{equation}
where the first term is the sum of four 1-body energies, the second
term is the sum of six 2-body interaction energies, the third term is
the sum of four 3-body interaction energies and the last term is the
4-body interaction energy. The first term is set to a constant value
which is the energy of total dissociation of N$_4$ to four N
atoms. Each 2-body term can be expressed by a 1D reproducing kernel
polynomial and corresponding RKHS PESs can be constructed from the
diatomic N$_2$ potential. The last two terms can be expressed by a
product of three and six 1D reproducing kernel polynomials. In this
work, the sum of the last two terms are calculated using RKHS
interpolation of the $E^{(3+4)}$ energies.  The sum of 3 and 4-body
interaction energies ($E^{(3+4)}$) is calculated as
\begin{equation}
  E^{(3+4)} =  V(r_{i}) -  \sum_{i=1}^4 V^{(1)}_i -  \sum_{i=1}^6 V^{(2)}_i(r_i).
\end{equation}
For all the cases the 1D kernel function ($k^{n,m}$) with smoothness
$n=2$ and asymptotic decay $m=6$ is used for the radial dimensions,
which is expressed as
\begin{equation}
 k^{2,6}(x,x') = \frac{1}{14}\frac{1}{x^7_{>}} - \frac{1}{18}\frac{x_<}{x^8_>},
 \label{eq:k_2_6}
\end{equation}
where, $x_>$ and $x_<$ are the larger and smaller values of $x$ and
$x'$, respectively, and the kernel smoothly decays to zero at long
range. Symmetry of the system can also be implemented within this
approach by considering all possible combinations for the 3 and 4-body
interaction energies. Here, it is worth to be mentioned that
interpolating the 3-body and 4-body terms separately should provide
more accurate energies, which is however not possible in this case as
the triatomic energies are not available.\\

\begin{table}[ht]
\caption{Root mean square error (RMSE), mean absolute error (MAE)
  computed the training data from the RKHS based PES in different
  energy ranges for N$_4$. Units of energies are in kcal/mol.}
\begin{tabular}{lccc|r}
\hline
\hline
   Energy range & Number of points & RMSE & MAE & RMSE\cite{TruhlarN4:2013} \\
    \hline
$E  \le 100.0$ & 678                         & 1.4     & 0.8    & 1.8   \\
$100.0 < E  \le 228.0$ & 1894        & 3.3     & 1.8    & 4.1    \\
$228.0 < E  \le 456.0$ & 11707      & 6.9     & 3.6    &  7.2   \\
$456.0 < E  \le 1000.0$ & 1608      & 16.5   & 9.6    & 18.0  \\
$1000.0 < E  < 1200.0 $ & 159      & 9.1     & 4.8    &           \\
\hline
\end{tabular}
\label{tab:errors}
\end{table}

\begin{figure}
\centering
\includegraphics[scale=1.8]{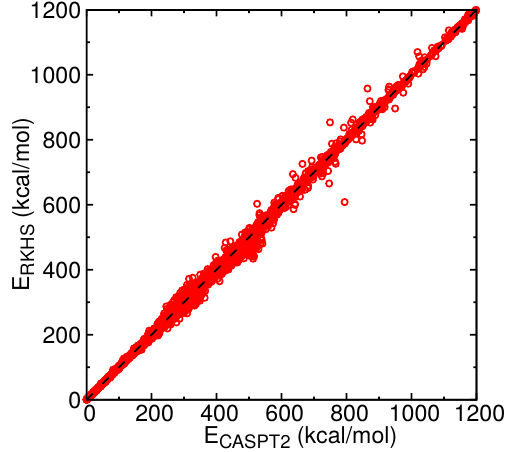}
\caption{Correlation between the RKHS and CASPT2 energies for 16063
  training data for N$_4$ system. The black dashed line shows ideal
  correlation between reference data and representation.}
\label{fig:n4corr}
\end{figure}

\begin{figure}
\centering
\includegraphics[scale=1.02]{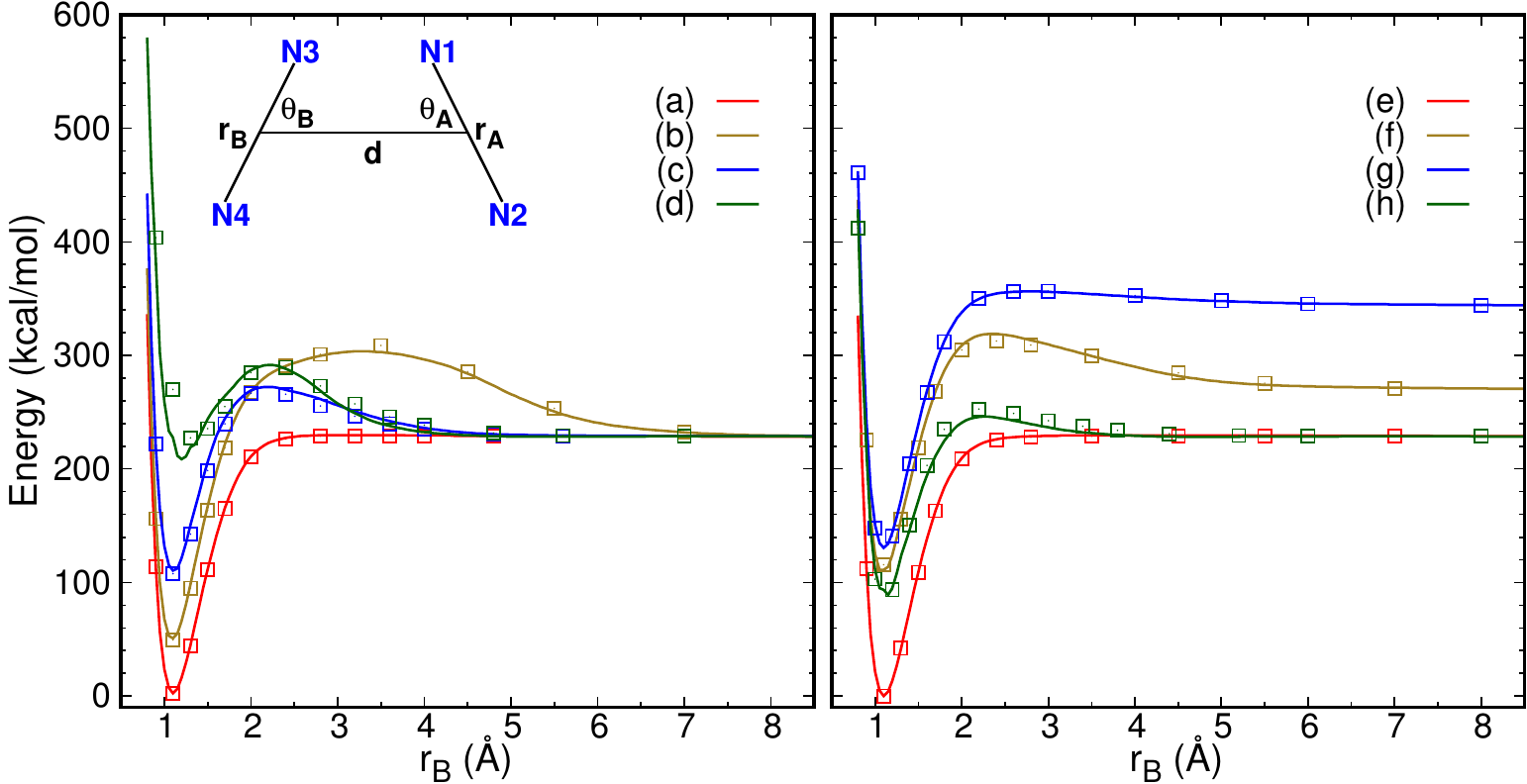}
\caption{Comparison between the test {\it ab initio} data (symbols)
  and RKHS interpolated energies (solid lines) for the dissociation
  curves N3-N4 (along $r_B$) for N$_2$+N$_2$ system with N1-N2 fixed
  at $r_A$. The angle between $r_A$ and $r_B$ is defined as
  $\phi$. (a) $r_A$ = 1.098 \AA , $d$ = 3.0 \AA , $\theta_A$ =
  $\theta_B$ = 90$^\circ$, $\phi$ = 0$^\circ$ (b) $r_A$ = 1.098 \AA ,
  $d$ = 2.4 \AA , $\theta_A$ = $\theta_B$ = 60$^\circ$, $\phi$ =
  0$^\circ$ (c) $r_A$ = 1.098 \AA , $d$ = 1.8 \AA , $\theta_A$ =
  $\theta_B$ = 90$^\circ$, $\phi$ = 90$^\circ$ (d) $r_A$ = 1.098 \AA ,
  $d$ = 2.0 \AA , $\theta_A$ = 0$^\circ$, $\theta_B$ = 90$^\circ$,
  $\phi$ = 0$^\circ$ (e) $r_A$ = 1.098 \AA , $d$ = 4.0 \AA ,
  $\theta_A$ = 120$^\circ$, $\theta_B$ = 60$^\circ$, $\phi$ =
  0$^\circ$ (f) $r_A$ = 1.298 \AA , $d$ = 2.6 \AA , $\theta_A$ =
  0$^\circ$, $\theta_B$ = 60$^\circ$, $\phi$ = 0$^\circ$ (g) $r_A$ =
  0.898 \AA , $d$ = 3.0 \AA , $\theta_A$ = 0$^\circ$, $\theta_B$ =
  60$^\circ$, $\phi$ = 0$^\circ$ (h) $r_A$ = 1.098 \AA , $d$ = 2.4 \AA
  , $\theta_A$ = 0$^\circ$, $\theta_B$ = 90$^\circ$, $\phi$ =
  0$^\circ$.}
\label{fig:n4comp}
\end{figure}

\noindent
The root mean square errors, mean absolute errors are computed for the
training data set and tabulated in Table \ref{tab:errors}.  The
correlation between the reference {\it ab initio} energies and RKHS
interpolated energies are plotted in Figure \ref{fig:n4corr} with an
$R^2 = 0.9981$. A few dissociation curves for the N$_2$ are plotted in
Figure \ref{fig:n4comp} for different configurations of the other
N$_2$ diatom. The {\it ab initio} energies shown in Figure
\ref{fig:n4comp} are not included in the RKHS training grid and show
that a RKHS can successfully reproduce the overall shape and values of
the unknown {\it ab initio} potential.\\

\noindent
Although techniques such as RKHS or permutationally invariant
polynomials can provide accurate representations, their extensions to
higher dimensions remains a challenge. Recently, the use of PIPs was
demonstrated for the PES of N-methyl acetamide which is an important
step in this direction.\cite{bowman.nma:2019} Additionally, the
(s)GDML approach\cite{chmiela2017machine,chmiela2019sgdml} has been
used to construct PESs for several small organic molecules, such as
ethanol, malondialdehyde and aspirin.\cite{sauceda2019construction}
Another challenge is to reduce the number of points required to define
such a PES. Efforts in this direction have recently shown that with as
few as 300 reference points the PES for scattering calculations in
OH+H$_2$ can be described from a fit based on Gaussian processes
together with Bayesian optimization.\cite{krems:2019} Nevertheless,
such high-accuracy representations of PESs for extended systems will
remain a challenge for both, the number of high-quality reference
calculations required and the type of inter- (and extra-)polation used
to represent them.\\

\begin{figure}
\includegraphics[width=0.9\textwidth]{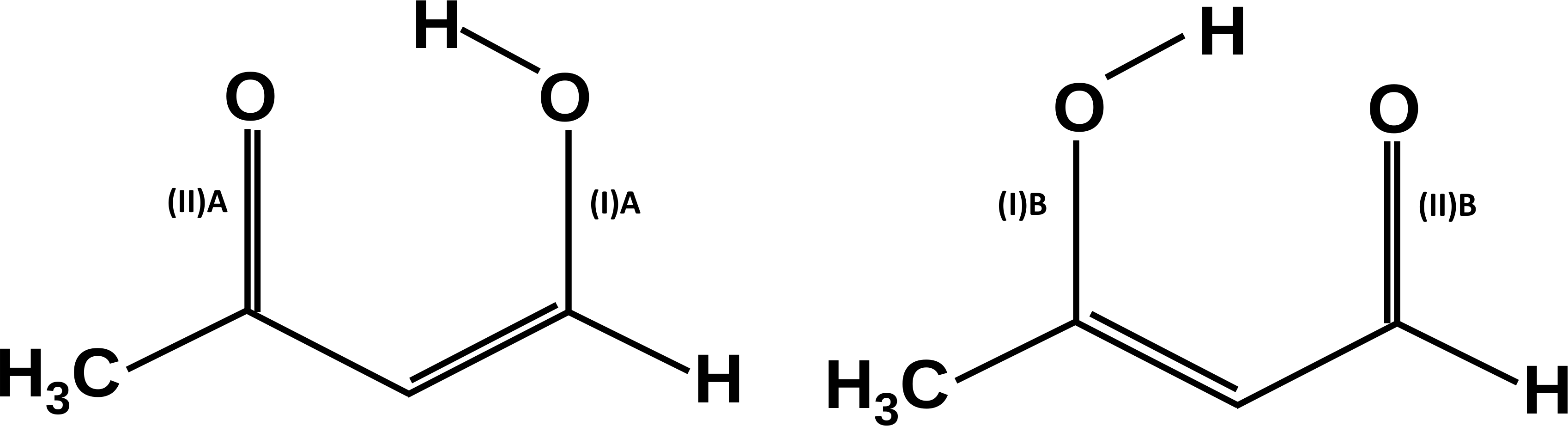}
\caption{The structure and local bonding motifs in singly methylated
  malonaldehyde. Depending on the position of the transferring
  hydrogen atom different single ((I)A and (I)B) and double ((II)A and
  (II)B) bonds arise. The distribution of the electrons will modify
  the stretching frequencies and therefore the force constants and
  equilibrium bond lengths. From optimizations at the MP2/6-31G(d,p)
  level the equilibrium separations of the single bonds (I)A and (I)B
  are 1.3224 \AA\/ and 1.3255 \AA\/ which compare with 1.2475 \AA\/
  and 1.2462 \AA\/ for the double bonds (II)A and (II)B,
  respectively.}
\label{fig:figpt}
\end{figure}

\noindent
Another important aspect of accurate studies of the energetics and
dynamics of molecular systems concerns the observation, that
``chemistry'' is often local. As an example, the details of a chemical
bond - its equilibrium separation and its strength - can depend
sensitively on the local environment which may play an important role
in applications such as infrared spectroscopy. As an example, singly
methylated malonaldehyde is considered. Depending on the position of
the proton, see Figure \ref{fig:figpt} the nature of the CO bond
changes. Overall, there are chemically 4 different CO bonds, two
single bonds (I)A and (I)B, and two double bonds (II)A and (II)B.  In
the language of an empirical force field, the equilibrium bond lengths
and the force constants change between these two structures in a
dynamical fashion, depending on the position of the transferring
hydrogen atom. Capturing such effects within an empirical force field
is possible, but laborious, as was recently done for the oxalate
anion.\cite{MM.oxa:2017}\\

\begin{figure}
\includegraphics[width=0.9\textwidth]{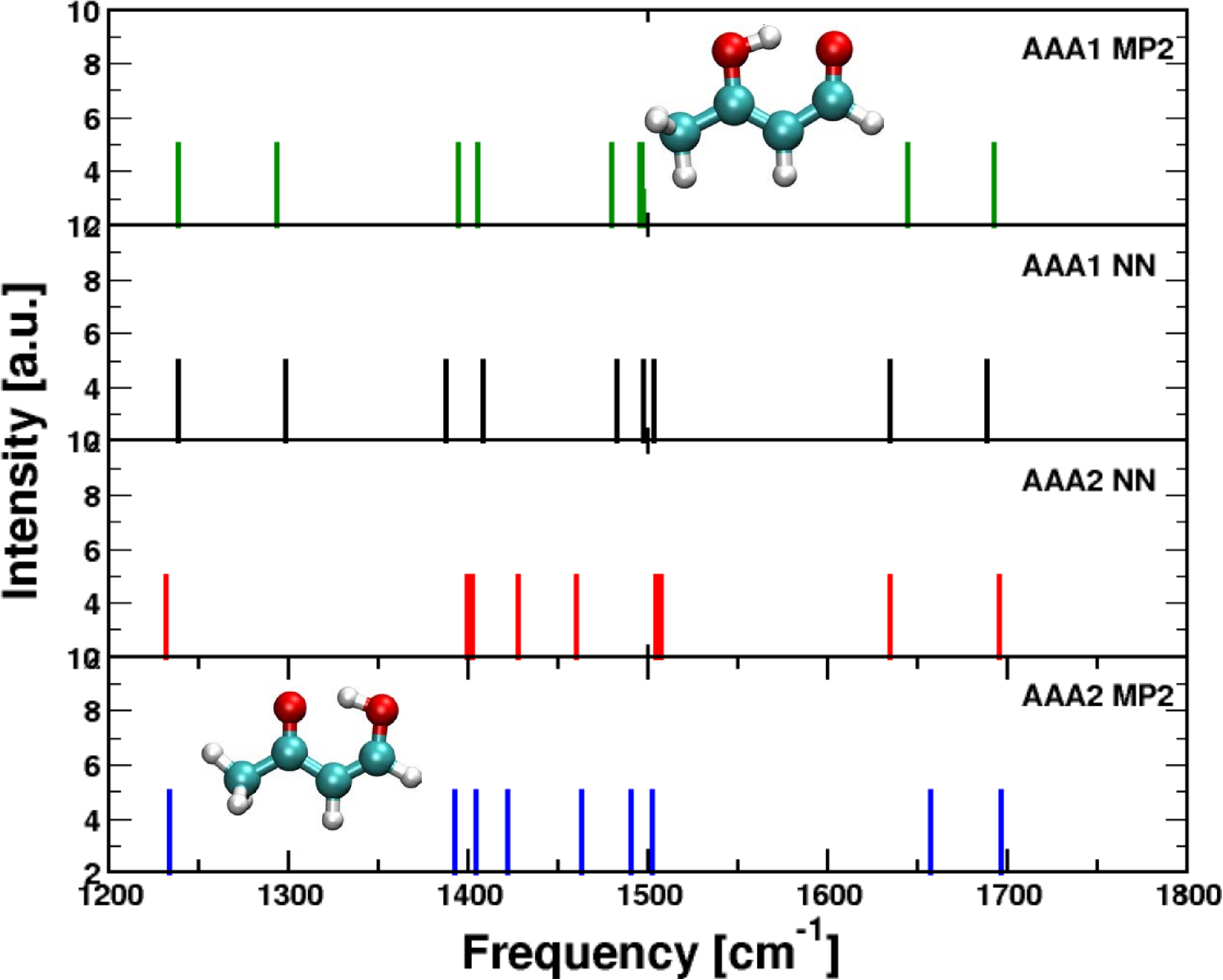}
\caption{The infrared spectrum of methylated malonaldehyde in the
  region of the CO stretch region. The bands at higher frequency
  (above 1600 cm$^{-1}$ are due to C=O bonds whereas those between
  1400 and 1500 cm$^{-1}$ involve a partial double bond for the CO
  stretch. The top and bottom panels are for normal modes from
  MP2/6-31G(d,p) calculations and the two middle panels from normal
  modes on the trained NN using PhysNet.}
\label{fig:mema}
\end{figure}

\noindent
Capturing such effects within a NN-trained global PES using PhysNet is
more convenient. As an example, the situation in singly-methylated
malonaldehyde (acetoacetaldehyde, AAA) is considered, see Figure
\ref{fig:figpt}. There are two CO motifs each of which can carry the
transferring hydrogen atom at the oxygen atom. Depending on whether
the hydrogen atom is on the OC-CH$_3$ or OC-C side the chemical nature
of the CO bond changes. This also influences the frequencies of the CO
stretch vibrations. Figure \ref{fig:mema} reports the infrared
spectrum from normal modes from MP2/6-311G(d,p) calculations and from
an NN trained on energies at the same level of theory. As is shown,
the normal modes from the electronic structure calculations from the
MP2/6-311G(d,p) for the two isomers (top and bottom panels) differ
appreciably in the range of the amide-I stretches. Above 1600
cm$^{-1}$ the harmonic frequencies occur at 1644 and 1692 cm$^{-1}$
for isomer AAA1 and at 1658 and 1696 cm$^{-1}$ for isomer AAA2. The NN
(middle two panels) is successful in capturing the higher frequency
(at 1689 and 1695 cm$^{-1}$ for the two isomers, respectively) whereas
for the lower frequency the two modes occur at 1635 and 1634
cm$^{-1}$. Additional modes involving CO stretch vibrations occur
between 1400 and 1500 cm$^{-1}$. Figure \ref{fig:mema} shows clear
differences for the patterns for AAA1 and AAA2 which are correctly
captured by the NN.\\

\noindent
In a conventional force field all these frequencies would be nearly
overlapping as the force field parameters for a CO bond does usually
not depend on whether a hydrogen is bonded to it or not. In order to
capture such an effect, the force field parameters for the CO bond
would need to depend on the bonding pattern of the molecule along the
dynamics trajectory. Encoding such detail into a conventional force
field is difficult and NN-trained PESs offer a natural way to do so.\\

\noindent
Another benefit yet to be explored that NN-trained PESs such as
PhysNet offer is the possibility to have fluctuating point charges for
a molecule without the need to explicitly parametrize the dependence
on the geometry. Modeling such effects within an empirical force field
is challenging.\cite{brooks:2004}\\

\noindent
A final challenge for high-dimensional PESs is including the chemical
environment, such as the effect of a solvent. Immersing a chemically
reacting system into an environment leads to pronounced changes. As an
example, double proton transfer in formic acid dimer in the gas phase
and in solution is considered. The parametrization used here was
adapted to yield the correct infrared spectrum in the gas
phase.\cite{MM.fad:2016} Recent high-resolution work has confirmed
that the barrier of 7.3 kcal/mol for the gas-phase PES is compatible
with the tunneling splitting observed in microwave
studies.\cite{fad:2019} Such a barrier height makes spontaneous
transitions rare. Hence, umbrella sampling simulations were combined
with the molecular mechanics with proton transfer (MMPT) force field
to determine the free energy barrier for DPT in the gas phase and in
solution. As a comparison, the simulations were also carried out by
using the Density-Functional Tight-Binding
(DFTB)\cite{sccdftb,Cui2014} method for the FAD. In both cases the
solvent was water represented as the TIP3P model.\cite{Jorg83}\\

\begin{figure}
\begin{center}
\includegraphics[width=0.85\textwidth]{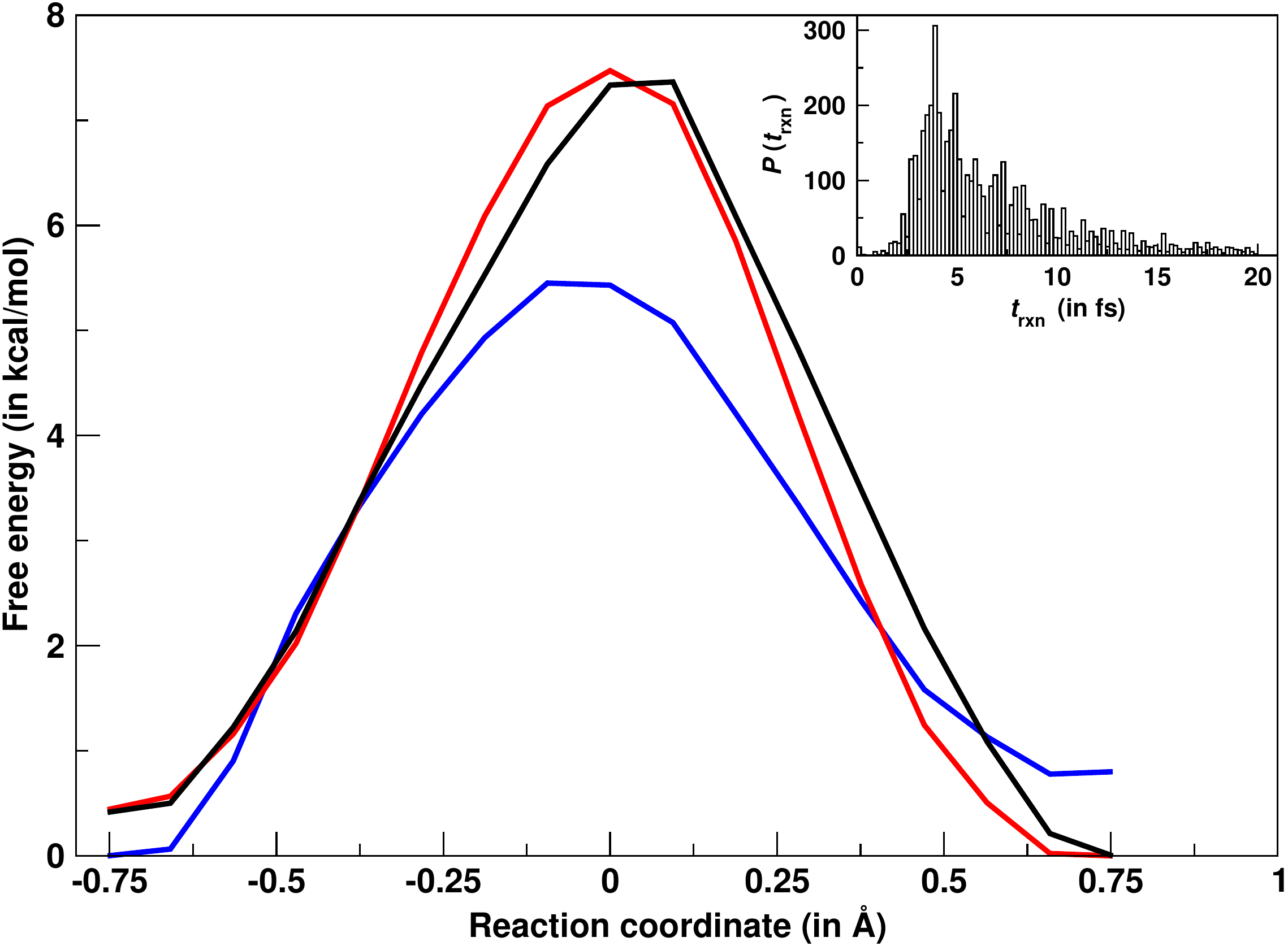}
\caption{Free energy as a function of reaction coordinate for proton
  transfer in gaseous and water-solvated FAD.  The blue and red curves
  show the free energy for FAD in the gas and solution phase
  respectively using the MMPT force field.  The energy profile in
  black is obtained for FAD in solution through DFTB treatment. In all
  cases, for the umbrella sampling procedure, 17 umbrella windows are
  located at 0.1 \AA\/ intervals and trajectories are propagated for
  50 ps. The probability distribution from different umbrellas are
  recombined using the weighted histogram analysis method
  (WHAM).\cite{Kumar.wham.1992}}
\label{fig:fad_free-ener}
\end{center}
\end{figure}

\noindent
The free energy barrier in the gas phase is $\Delta G = 5.4$ kcal/mol
which increases to 7.5 kcal/mol in water, see
Fig.~\ref{fig:fad_free-ener}. With DFTB3 the barrier height in
solution is similar (7.3 kcal/mol) to that with the MMPT
parametrization. In all cases, FAD undergoes a concerted double proton
transfer to interconvert between two equivalent forms resulting in a
symmetric potential. The nature of the transition state was verified
by running 5000 structures from the umbrella sampling simulations at
the TS, starting with zero velocity, and propagating them for 1 ps in
an $NVE$ ensemble. The fraction of reactants and products obtained are
0.54 and 0.46, indicating that the configurations sampled in the
umbrella sampling simulations indeed correspond to a transition state
and lie midway between reactants and products and are equally likely
to relax into either stable state.\\

\begin{figure}
\begin{center}
\includegraphics[width=0.85\textwidth]{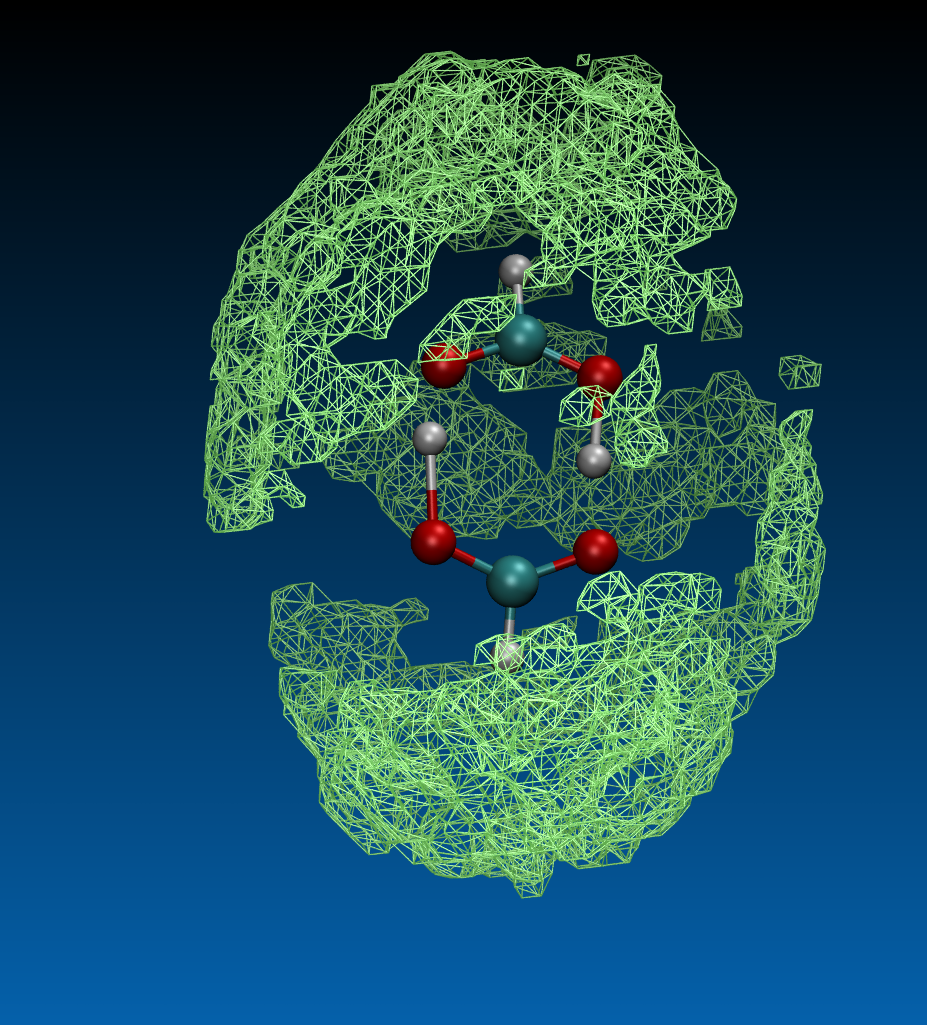}
\caption{Solvent distribution around FAD for the transition state
  ensemble from 5000 transition states sampled from umbrella sampling
  simulations.}
\label{fig:fad_solv}
\end{center}
\end{figure}

\noindent
From these simulations it is also possible to determine the time to
product or reactant which is reported in the the inset of Figure
\ref{fig:fad_free-ener}. The most probable time is $\sim 5$ fs with a
wide distribution extending out to to 20 fs. This is typical for a
waiting time distribution and indicates that multiple degrees of
freedom are involved.\\

\noindent
The methods discussed in the present work have all their advantages
and shortcomings. Depending on the application at hand the methods
provide different efficiencies and accuracies and are more or less
straightforward to apply. In the following, the three approaches
discussed here are compared by looking at them from different
perspectives.
\begin{itemize}
\item For small gas phase systems such as tri- and tetraatomics,
  RKHSs, PIPs and NN-based force fields are powerful methods for
  accurate investigations of their reactive dynamics. Empirical force
  fields are clearly not intended and suitable for this.
\item For medium-sized molecules (up to $\sim 10$ atoms) in the gas
  phase, reactive MD methods, such as EVB\cite{EVB-Warshel1980} (not
  explicitly discussed here or multi state reactive MD), NNs, or
  suitably parametrized force fields (polarizable or non-polarizable)
  including multipoles are viable representations. PIPs or RKHSs will
  eventually become cumbersome to parametrize and computationally
  expensive to evaluate.
\item Systems with $\sim 10$ atoms in solution can be described by
  refined FFs and reactive MD simulations. NNs, such as Physnet, would
  be a very attractive possibility, as they include fluctuating
  charges by construction. Also, capturing changes in the bond
  character depending on the chemical environment (see discussion of
  methylated MA above) is readily possible. However, an open technical
  question is how to include the effect of the environment in training
  the NN.
\item Finally, for macromolecules in solution, such as proteins,
  either refined reactive FFs or a combination of RKHS and a FF has
  shown to provide meaningful ways to extend quantitative, reactive
  simjlations to condensed phase systems. Extending such approaches,
  akin to mixed QM/MM simulations but treating the reactive part with
  a NN, may provide even better accuracy.
\end{itemize}

\noindent
Multidimensional PESs are a powerful way to run high-quality atomistic
simulations for gas- and condensed phase systems. Recent progress
concerns the accurate, routine representation of PESs based on RKHSs
or PIPs. As an exciting alternative, NN-based PESs have also become
available. Despite this progress, extension of these techniques to
simulations in solution and multiple dimensions remain a
challenge. Attractive future possibilities are simulations which
capture the changes in local chemistry or in the atomic charges
without the need to explicitly parametrize them as a function of
geometry. This is possible with approaches as those used in PhysNet.\\

\section*{Acknowledgments}
The authors acknowledge financial support from the Swiss National
Science Foundation (NCCR-MUST and Grant No. 200021-7117810) and the
University of Basel.

\bibliography{references}

\end{document}